\documentclass[10pt,lettersize,journal]{IEEEtran}
\usepackage{amsmath,amsfonts}
\usepackage{algorithmic}
\usepackage{algorithm}
\usepackage{array}
\usepackage[caption=false,font=scriptsize,labelfont=rm,textfont=rm]{subfig}
\usepackage{textcomp}
\usepackage{stfloats}
\usepackage{url}
\usepackage{verbatim}
\usepackage{graphicx}
\usepackage{cite}
\usepackage{tcolorbox}
\usepackage{amssymb}
\usepackage{graphicx}
\usepackage{relsize}
\usepackage[colorinlistoftodos]{todonotes}
\usepackage{hyperref}
\usepackage{pdfpages}
\usepackage{lipsum}
\usepackage{lmodern}
\usepackage{tcolorbox}
\usepackage{blindtext}
\usepackage[T1]{fontenc}
\usepackage{cuted} 
\usepackage{multirow}
\usepackage{xcolor}
\usepackage{comment}
\usepackage{booktabs}
\usepackage{times} 
\usepackage{soul}

\graphicspath{{Figures_v4/}}

\begin{document}

\title{Quantization of KLT Matrices via GMRF Modeling of Image Blocks for Adaptive Transform Coding}

\author{%
Rashmi Boragolla, and Pradeepa Yahampath\\[0.5em]
{\small\begin{minipage}{\linewidth}\begin{center}
\begin{tabular}{c}
Department of Electrical and Computer Engineering, University of Manitoba \\
75 Chancellors Circle, Winnipeg, MB, R3T 2N2, Canada \\
\url{borbnwmr@myumanitoba.ca}, \url{pradeepa.yahampath@umanitoba.ca} 
\end{tabular}
\end{center}\end{minipage}}
}

\maketitle
\thispagestyle{empty}

\begin{abstract}
Forward adaptive transform coding of images requires a codebook of transform matrices from which the best transform can be chosen for each macroblock. Codebook construction is a problem of designing a quantizer for Karhunen-L\'{o}eve transform (KLT) matrices estimated from sample image blocks. We present a novel method for KLT matrix quantization based on a  finite-lattice non-causal homogeneous Gauss-Markov random field (GMRF) model with asymmetric Neumann boundary conditions for blocks in natural images. The matrix quantization problem is solved in the GMRF parameter space, simplifying the harder problem of quantizing a large matrix subject to an orthonormality constraint to a low-dimensional vector quantization problem. Typically used GMRF parameter estimation methods such as maximum-likelihood (ML) do not necessarily maximize the coding performance of the resulting transform matrices.  To this end we propose a method for GMRF parameter estimation from sample image data, which maximizes the high-rate transform coding gain. We also investigate the application of GMRF-based transforms to variable block-size adaptive transform coding.
\end{abstract}

\section{Introduction}

Image and video coding algorithms rely on transform coding for efficient block-wise compression of spatial data. In practice, the two-dimensional discrete cosine transform (2D-DCT) has been found to be the most effective transform for non-adaptive coding. Under the mean square error (MSE) criterion, the KLT is the optimal transform for coding stationary Gaussian sources \cite{Gersho}. Indeed, for a certain class of GMRFs the 2D-DCT is a KLT \cite{Moura2}. However, not all textures appearing in natural images conform to this special class of GMRFs. Furthermore, due to the non-stationary nature of natural images and video, the optimal transform tends to vary depend on the spatial location. Consequently, {\em content-adaptive} transforms (CAT) can often outperform the 2D-DCT \cite{Zhang,Zhao1}.

In general, adaptive coding can be either forward or backward adaptive \cite{Gersho}, with the former being more widely used in image and video coding. There are two general approaches to forward CAT coding. The first is the  use of an assortment of standard trigonometric transforms \cite{Zhao,Zhao1}. The second involves estimating the local KLT for each coded image block and using a quantized version of the estimated KLT for transform coding. For each coded block, the quantization index of the transform is signaled to the decoder, alongside the transform coefficients. In practice KLT quantization is realized by off-line learning of a {\em codebook} of transform matrices, from which the best transform for each coded block (or a macroblock) is chosen \cite{Zeng,Zhao3,Zou,Zhang}.

Learning a CAT codebook from sample data involves estimating a finite set of orthonormal matrices which is optimal in some sense for transform coding an ensemble of image blocks (training set.) In other words, one has to design a quantizer for a random orthonormal matrix whose empirical distribution is defined by KLTs of sample image blocks. One recently proposed general method for learning optimal transforms from data is {\em sparse orthonormal transforms} \cite{Sezer2}, which is based on a pursuit-type algorithm. Another general approach is \cite{Boragolla1} which performs quantization directly on the manifold of orthonormal matrices using an iterative algorithm resembling the generalized Lloyd algorithm. However, these {\em non-parametric} methods do have disadvantages. First, such methods carry out matrix optimization in a very high dimensional space, and hence are susceptible to the curse of dimensionality. For example, a non-separable transform for an $N \times N$ image block is a $K \times K$ orthonormal matrix $\pmb{T}$ where $K=N^2$. The orthonormality constraint implies that the space of $\pmb{T}$ is the $K(K-1)/2$ dimensional Euclidean space. Thus, even for $8 \times 8$ image blocks $\pmb{T}$ lies in a 2016-dimensional space. The second shortcoming of the aforementioned methods is that the transform matrices are not {\em scalable.} If multiple transform block sizes are to be used, such as in video compression \cite{Zhao1} then a separate matrix codebook must be designed for each block size. It is useful to have a basic design from which transforms of different sizes can be derived, as with the DCT.

In this paper, we propose a novel model-based approach to transform matrix codebook design which can potentially address both dimensionality and scalability issues\footnote{This work was presented in part at DCC 2024 \cite{Boragolla2}.}. The key idea is to parameterize the  $K \times K$ KLT  of an image-block by modeling it as a finite-lattice non-causal homogeneous GMRF whose precision matrix is completely defined by a number of parameters  much smaller than the dimension of the KLT,  $K(K-1)/2$. Since the KLT of a GMRF is the eigenvector matrix of the precision matrix \cite{Zhang1}, this parameterization maps the KLT to a low dimensional Euclidean space, simplifying our transform matrix codebook design problem of quantizing a large random matrix subject to an orthonormality constraint to one of quantizing a low-dimensional GMRF parameter vector. In fact, it is our observation that, for textures in natural images, a 2nd-order GMRF with 4 parameters is sufficient, implying that KLT of an image block with a given texture can be represented by only 4-parameters, regardless of the block-size. In this paper we will refer to a transform matrix constructed through GMRF parameterization as a {\em GMRF-based transform} (GMRFT.) A codebook of GMRFT matrices is designed by first estimating a population of GMRF parameter vectors from a suitable training set of image blocks and vector quantizing the parameter vectors in that population. 

GMRFTs are inherently scalable in size, since a given GMRF parameter vector can represent an image block of any size so long as the model is appropriate for the image texture. Essentially, our transform selection in  an adaptive coding setting can be viewed as one of choosing the best texture model for a given image block  from a codebook of models such that the transform coding error is minimized. As such, we also propose here a new transform coding optimized GMRF parameter estimation method, which according to our experimental results, is superior to other commonly used generic parameter estimation methods such as ML method, when the end goal is to estimate a GMRF precision matrix whose eigenvector matrix produces a high transform coding gain for image blocks.

GMRFTs are not restricted to non-separable transforms. The 2D-DCT, which is a separable transform, is the KLT for a finite-lattice non-causal homogeneous GMRF model with a diagonally symmetric neighborhood structure \cite{Moura2}. The model proposed in this paper reduces to the latter structure for a certain choice of model parameters and hence the 2D-DCT is simply a special case of a broader class of GMRFTs studied in this paper. Numerical results obtained with an extensive set of experiments showed that the proposed  GMRFTs are superior to the 2D-DCT for many image blocks in natural images. It was also observed that, despite being a low-dimensional approximation to the KLT, GMRFTs are competitive with recently reported model-free (non-parametric) data driven transforms \cite{Sezer2,Boragolla1} in fixed block-size coding. Furthermore, when used for quad-tree driven variable block-size coding, GMRFTs not only outperformed these fixed block-size transforms, but also the variable block-size counterparts based on the  2D-DCT and its variant the directional DCT (DDCT) \cite{Zeng}.

\subsubsection*{Related work}
GMRF models have been previously used in various ways for image and video compression. Early work focused on using GMRFs for coding image textures directly in spatial domain  \cite{Balle,Balram1,Chellappa}. More recent and closely related to ours is the approach of using GMRF models to estimate graph Fourier transforms for image data \cite{Egilmez1,Egilmez4,Fracastoro1}. In this case the GMRF model is constrained to be {\em attractive} (only positive correlations among pixels allowed) and  a graph-Laplacian structure is imposed on its precision matrix \cite{Egilmez2}. Our approach does not require such constraints. The work in \cite{Egilmez1,Egilmez4} focuses on compression of prediction residuals in a HEVC video codec using graphs derived from 1D Gauss-Markov process models (line graphs) with positive correlations. In \cite{Fracastoro1} an attractive GMRF model with an unstructured  precision matrix (without assuming a homogeneous neighborhood structure and any parameterization as we do) is used to derive graph Fourier transforms for image compression. Our experimental results for natural images show that transforms derived from attractive GMRF models are inferior (in terms of coding gain) to those derived from  unconstrained models as considered in this paper.   

The remainder of the paper is organized as follows. Section \ref{sec_gmrf} introduces finite lattice homogeneous GMRF models. Section \ref{sec_model} then presents the GMRF model upon which the proposed GMRFTs are based and the GMRFT matrix codebook design procedure. Variable block-size (scalable) transform matrix construction using GMRF parameterization is discussed in Sec. \ref{sec_scalable_trans}. Experimental results are presented and discussed in Sec. \ref{sec_results}. Concluding remarks appear in Sec. \ref{sec_conclude}.

\section{Finite-lattice homogeneous GMRF models} \label{sec_gmrf}
\begin{figure}[!tb]
  \centering
  \subfloat[]{\includegraphics[width=0.2\textwidth]{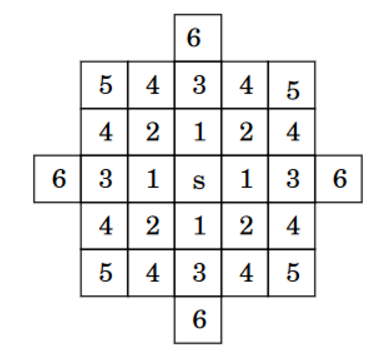}}\\ 
  \subfloat[]{\includegraphics[width=0.225\textwidth]{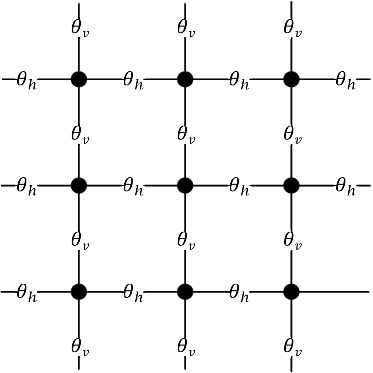}}\hspace{1ex}
  \subfloat[]{\includegraphics[width=0.225\textwidth]{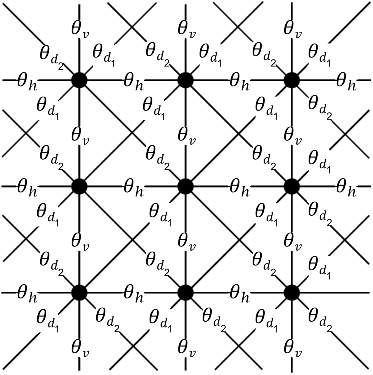}}
\caption{ \label{fig:GMRF_NH} (a) Non-causal GMRF neighborhood structures of orders 1 to 6 for the pixel $s$. All pixels labeled by values $t \leq M$ belong to the M-th order neighborhood of $s$. (b) 1st-order and (c) 2nd order GMRF models with homogeneous spatial interactions.}
\end{figure}

We will model an image block ${\pmb U} \in {\mathbb R}^{N \times N}$ by a 2D {\em non-causal} and {\em homogeneous} GMRF defined on a finite $N \times N$ lattice, see \cite{Rue1,Chellappa3,Moura4,Balram1} for detailed descriptions. An $N \times N$ finite-lattice GMRF $\pmb{U}$ is a set of jointly Gaussian random variables $\{ U_{l,m}\}$, $l \in \{ 1\ldots,N \}$ and $m \in \{ 1\ldots,N \}$, arranged on a  2D plane such that each $U_{l,m}$ is conditionally independent of all other variables, given the neighbor-set ({\em neighborhood}) ${\mathcal N}_{l,m}$ of $U_{l,m}$, i.e. $p(U_{l,m}|U_{s,t}\in\pmb{U}, s \neq l, t \neq m)= p(U_{l,m}|{\mathcal N}_{l,m})$ (Markov property.) The GMRF is {\em homogeneous} if the neighborhood structure and the conditional pdfs  $p(U_{l,m}|{\mathcal N}_{l,m})$ are invariant with respect to the  spatial location of $U_{l,m}$ in the lattice. In addition, a finite lattice non-causal homogeneous GMRF also requires the specification of boundary conditions which are also spatially invariant with respect to all boundary locations \cite{Moura4,Balram1}. The size of the neighborhood is referred to as the {\em order} of the GMRF. Fig. \ref{fig:GMRF_NH}(a) shows possible neighborhood structures of orders 1 to 6 with respect to the pixel $s$, where the numbers indicate the model order. For example, all pixels labeled by values $\leq 2$ belong to the 2nd-order neighborhood of $s$. An important property of a GMRF is that the Markov property implied by the neighborhood structure is encoded in its {\em precision matrix} \cite{Rue1}. 

Let the vector form of the GMRF be $\pmb{X}=\mbox{vec}(\pmb{U})$ where $\pmb{X}= (X_1,\:X_2,\:\cdots,\:X_K)^T$ and $K =N^2$. Let the mean vector $E [ \pmb{X}]=\pmb{\mu}$ and the covariance matrix $E [ (\pmb{X}-\pmb{\mu})(\pmb{X}-\pmb{\mu})^T]=\pmb{C}$, which is a $K\times K$  symmetric positive definite (s.p.d.) matrix. The density function of  $\pmb{X}$ is given by
\begin{align}
p(\pmb{X}) = \frac{\sqrt{\vert\pmb{Q}\vert}}{(2\pi)^{K/2}} \exp\left( -\frac{1}{2}\left(\pmb{X} - \pmb{\mu}\right)^T\pmb{Q} \left(\pmb{X} - \pmb{\mu}\right) \right), \label{gmrf_model}
\end{align}
where $\pmb{Q}=\pmb{C}^{-1}$ is the precision matrix \cite[Theorem 2.3]{Rue1} which is also s.p.d. If $\pmb{X}$ is a GMRF with respect to a given neighborhood structure, then  $Q_{i,j} \neq 0$ for $i \neq j$, if and only if $X_i$ and $X_j$ are mutually neighbors. The $i$-th diagonal element $Q_{i,i}$ of the  precision matrix is the conditional precision (inverse of variance) of $X_i$  given all other variables in $\pmb{X}$. Let ${\mathcal N}_i$ denote the neighbor-set of $X_i$. The conditional means are then given by
\begin{align} \label{cond_mean}
  E\left[ X_i|{\mathcal N}_i\right]= \mu_i-\sum_{X_j \in {\mathcal N}_i}\frac{Q_{i,j}}{Q_{i,i}}(X_j -\mu_j).
\end{align}
The off-diagonal elements $Q_{i,j}$, $ i \neq j$ determine the conditional correlation coefficient of $X_i$ and $X_j$, given the rest of the variables in $\pmb{X}$. This implies that if the order of the GMRF is small, $\pmb{Q}$ is a highly sparse matrix. In other words, as the eigenvectors of $\pmb{Q}$ and $\pmb{C}$ are identical, the KLT of an image block modeled by a homogeneous GMRF can be represented by a small number of parameters which depends only on the neighborhood-order. Since  the dependencies between pixels in natural images tend to be local, there is very little to be gained by using neighborhoods larger than the  6-th order neighborhood shown Fig. \ref{fig:GMRF_NH}(a).

\subsubsection*{Parameterization of a non-causal homogeneous GMRF}
Consider a mean-zero infinite-lattice non-causal homogeneous GMRF, which can be specified by a 2D non-causal auto-regressive (AR) representation \cite{Woods}
\begin{align} \label{ar1}
 U_{l,m}=\sum_{s \neq 0, t \neq 0}\theta_{s,t}U_{l-s,m-t}+\epsilon_{l,m},
\end{align}
$l=1,2,\ldots,$ and $m=1,2,\ldots$, where $\{ \epsilon_{l,m} \}$ is a mean-zero homogeneous GMRF such that $E[ U_{l-s,m-t}\epsilon_{l,m}]=0$ for $s \neq 0$ and $t \neq 0$. Spatially invariant 2D AR coefficients $\theta_{s,t}$ in (\ref{ar1}), which are referred to as {\em spatial interactions} or {\em field potentials}, are non-zero only if $U_{l,m}$ and $U_{l-s,m-t}$ are  neighbors. Now suppose we obtain a finite-lattice homogeneous GMRF by truncating the infinite-lattice  homogeneous GMRF and applying suitable boundary conditions \cite{Moura4}. Given the boundary conditions, the non-causal AR representation of the resulting finite-lattice GMRF $\pmb{U} \in {\mathbb R}^{N \times N}$ can be derived from (\ref{ar1}) [see Sec. \ref{sec_model}]. The AR representation of $\pmb{X}=\mbox{vec}(\pmb{U})$ has the form
\begin{align} \label{ar2}
 \beta_{i,i}X_{i}=\sum_{j \neq i}\beta_{i,j}X_j+W_{i}, \quad i=1,\dots,K,
\end{align}
 where $\pmb{W}=(W_1,\ldots, W_K)^T$ is a mean-zero  Gaussian vector and the coefficients $\{ \beta_{.,.} \}$ are functions of $\{ \theta_{.,.}\}$ and depend on the boundary conditions. From (\ref{cond_mean}) it directly follows that, for jointly Gaussian $\pmb{X}$ with $\pmb{\mu}=\pmb{0}$, we can write
\begin{align} \label{ar3}
  Q_{i,i}X_i = -\sum_{X_j \in {\mathcal N}_i} Q_{i,j}X_j+V_i, \quad i=1,\ldots,K,
\end{align}
where $\pmb{V}=(V_1,\ldots,V_K)^T$ is a mean-zero Gaussian vector. Representations (\ref{ar2}) and (\ref{ar3}) are equivalent up to a scaling factor provided $\{ \theta_{s,t}\}$ are such that $\beta_{i,j}=\beta_{j,i}$. That is, 
\begin{align*}
  Q_{i,j}=Q_{j,i}= \left \{
  \begin{array}{ll}
    c\beta_{i,j} & i=j \\
    -c\beta_{i,j} & \mbox{ otherwise,}
  \end{array}
  \right.
\end{align*}
where $c>0$ is some constant. This constant can be ignored as scaling $\pmb{Q}$ does not affect its eigenvectors (KLT of $\pmb{X}$.) We will therefore denote the precision matrix by $\pmb{Q}(\pmb{\theta})$ where $\pmb{\theta}$ is the vector of non-zero spatial interactions $\{ \theta_{s,t} \}$ in (\ref{ar1}) which we will refer to as {\em GMRF parameters.}

\section{Proposed GMRFT: Modeling and design } \label{sec_model}
\subsection{Image model and motivation} \label{pro_model}
For the purpose of estimating a KLT for transform coding, we model each image block as a realization of some  finite lattice non-causal homogeneous GMRF with a parameter vector $\pmb{\theta} \in {\mathbb R}^p$.  Examples of infinite lattice non-causal homogeneous GMRFs of orders $1$ and $2$ are shown in Figs. \ref{fig:GMRF_NH}(b) and \ref{fig:GMRF_NH}(c) where $\pmb{\theta}=(\theta_h,\theta_v)^T$ and $\pmb{\theta}=(\theta_h,\theta_v,\theta_{d_1},\theta_{d_2})^T$ respectively. A finite-lattice homogeneous GMRF can be obtained by truncating an infinite lattice homogeneous GMRF and applying suitable boundary conditions so that the neighborhoods of all pixels in the finite lattice have the same order $p$. The choice of boundary conditions impacts the eigen structure of its precision matrix $\pmb{Q}$. For example, in the 1D case, {\em periodic boundary conditions} result in a finite lattice non-causal homogeneous GMRF whose ${\pmb Q}$ matrix is a circulant matrix, and hence the KLT is the basis vectors of the {\em discrete Fourier transform} (DFT) \cite{Rue1}. On the other hand with {\em asymmetric Neumann boundary conditions}, $\pmb{Q}$ matrix has a structure such that the KLT is the basis vectors of the DCT \cite{Moura2}. In the 2D case, periodic boundary conditions result in a finite lattice non-causal homogeneous GMRF whose $\pmb{Q}$ matrix is block circulant and hence the KLT is the 2D-DFT \cite{Rue1}. Other results generalize to the 2D case under certain symmetry conditions on the neighborhood structure. In particular, when the spatial interactions of a non-causal homogeneous 2D GMRF are {\em diagonally symmetric}, asymmetric Neumann boundary conditions result in a finite-lattice GMRF whose KLT is the 2D-DCT \cite{Moura2}.

{\em Diagonally symmetric fields}: Consider the lattice variable $U_{l,m}$ in a non-causal homogeneous  GMRF, and let $\theta_{s,t}$ be the spatial interactions between $U_{l,m}$ and its neighbor  $U_{l-s,m-t}$. The GMRF is diagonally symmetric if
\begin{align} \label{diag_sym}
  \theta_{s,t}=\theta_{-s,t}=\theta_{s,-t}=\theta_{-s,-t}.
\end{align}
For example, the 1st-order field in Fig. \ref{fig:GMRF_NH}(b) is diagonally symmetric. The 2nd-order field in Fig. \ref{fig:GMRF_NH}(c) is diagonally symmetric only if $\theta_{d_1}=\theta_{d_2}$.

\begin{figure}[!tb]
  \centering
  \includegraphics[width=0.3\textwidth]{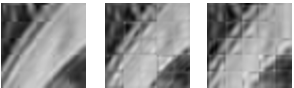}
  \caption{\label{lena_hat} Effect of boundary conditions on $8 \times 8$ blocks (taken from {\em Lena} image): assymetric Neumann (left), periodic (center), and Dirichlet.}
\end{figure}

In general, energy packing efficiency \cite{Clarke} of the 2D-DFT is poorer compared to the 2D-DCT due to the difference in respective boundary conditions - periodic boundary conditions can introduce sharp transitions whereas the asymmetric Neumann boundary conditions result in smooth transitions \cite{Clarke}. Asymmetric Neumann boundary conditions assign values to pixels outside an image block using those inside the block such that the image intensity gradient (backward difference) normal to the block boundary is zero \cite{Moura4,Ng}. Fig. \ref{lena_hat} compares the effects of commonly used boundary conditions on GMRFTs. These images have been reconstructed by considering only  12.5\% of the transform coefficients for each $8 \times 8$ pixel block. Note the artifacts due to abrupt transitions forced by periodic and Dirichlet (zero) boundary conditions \cite{Rue1}.

\begin{figure}[!tb]
  \centering
  \subfloat[]{\includegraphics[width=0.35\textwidth]{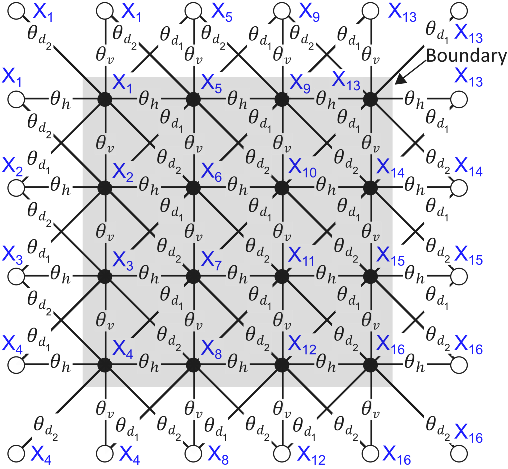}}\\
  \subfloat[]{\includegraphics[width=0.35\textwidth]{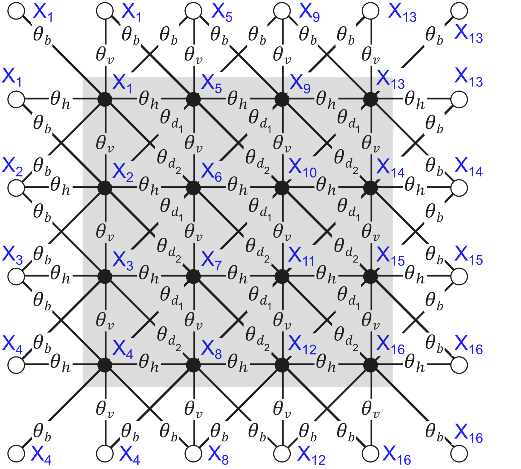}}
\caption{ \label{order2_model} (a) A finite lattice homogeneous field (gray square) satisfying conditions C1 and C2.  (b) modified field with diagonally symmetric spatial interactions along the boundary such that precision matrix is symmetric.}
\end{figure}

Based on above observations, in order to model image blocks whose KLT is more general than the 2D-DCT, we propose to use a finite lattice non-causal homogeneous GMRF  model which satisfies the following two conditions.
\begin{itemize}
\item[]{\em C1:} Spatial interactions  must  {\em not} be restricted by the diagonal symmetry condition (\ref{diag_sym}). 
\item[]{\em C2:} Asymmetric Neumann boundary conditions must be enforced. 
\end{itemize}
Fig. \ref{order2_model}(a) shows a $4 \times 4$ finite lattice satisfying these conditions, which is obtained by applying asymmetric Neumann boundary conditions to the 2nd-order infinite lattice in Fig. \ref{fig:GMRF_NH}(c).

It can however be verified that the field in Fig. \ref{order2_model}(a) does not satisfy the required symmetry condition $\beta_{i,j}=\beta_{j,i}$ in (\ref{ar2}) for boundary lattice points. In general, it can be shown that imposing asymmetric Neumann boundary conditions on a homogeneous field of order 2 or higher results in a valid AR representation (\ref{ar2}) only if the diagonal symmetry is satisfied for spatial interaction between the points inside and outside the lattice. In order to resolve this issue, we let $\theta_{d_1}=\theta_{d_2}=\theta_{b}$ outside the boundary as shown in Fig. \ref{order2_model}(b). This model has 5 parameters $(\theta_v,\theta_h, \theta_{d_1}, \theta_{d_2}, \theta_{b})$.  The precision matrix of this model, which is shown in Fig. \ref{eq_Q_mat_4x4}, can be obtained by applying (\ref{ar2}) to every lattice point and  comparing the result with (\ref{ar3}). Our experimental results showed that replacing the diagonal interactions outside the boundary by the average of the corresponding diagonal interactions inside the lattice did not result in a noticeable difference in the coding performance of the resulting transform matrices for image block-sizes $4 \times 4$ and larger, regardless of the model order. This eliminates the need to define an additional parameter. For example, in the 2-nd order model in Fig. \ref{order2_model}(b), we can let $\theta_{b}=\frac{\theta_{d_1}+\theta_{d_2}}{2}$,  
\subsubsection*{Remarks}
\begin{enumerate}
\item Diagonal symmetry ($\theta_{d_1}=\theta_{d_2}$) is a special case of the above described model and  hence the 2D-DCT is included in the class of transforms defined by this model.
\item A 1st-order model [e.g., Fig. \ref{fig:GMRF_NH}(b)] is always diagonally symmetric and hence only 2nd or higher-order models are useful for improving on the 2D-DCT.
\end{enumerate}

\begin{figure*}[!tb]
\centering
  \begin{align*}
\left[
\begin{smallmatrix}
q_1 & q_4 & 0   & 0   & q_5 & -\theta_{d_2} & 0   & 0   & 0   & 0   & 0   & 0   & 0   & 0   & 0   & 0 \\
q_4 & q_2 & q_4 & 0   & -\theta_{d_1} & -\theta_h   & -\theta_{d_2} & 0   & 0   & 0   & 0   & 0   & 0   & 0   & 0   & 0 \\
0   & q_4 & q_2 & q_4 & 0   & -\theta_{d_1} & -\theta_h   & -\theta_{d_2} & 0   & 0   & 0   & 0   & 0   & 0   & 0   & 0  \\
0   & 0   & q_4 & q_1 & 0   & 0   & -\theta_{d_1} & q_5 & 0   & 0   & 0   & 0   & 0   & 0   & 0   & 0 \\
q_5 & -\theta_{d_1} & 0   & 0   & q_3 & -\theta_v   & 0   & 0   & q_5 & -\theta_{d_2} & 0   & 0   & 0   & 0   & 0   & 0\\
-\theta_{d_2} & -\theta_h   & -\theta_{d_1} & 0   & -\theta_v   & 1   & -\theta_v   & 0   & -\theta_{d_1} & -\theta_{h} & -\theta_{d_2}   & 0   & 0   & 0   & 0   & 0\\
0   & -\theta_{d_2} & -\theta_h   & -\theta_{d_1} & 0   & -\theta_h   & 1   & -\theta_h   & 0   & -\theta_{d_1} & -\theta_h   & -\theta_{d_2} & 0   & 0   & 0   & 0\\
0   & 0   & -\theta_{d_2} & q_5 & 0   & 0   & -\theta_v   & q_3 & 0   & 0   & -\theta_{d_1} & q_5 & 0   & 0   & 0   & 0\\
0   & 0   & 0   & 0   & q_5 & -\theta_{d_1} & 0   & 0   & q_3 & -\theta_v   & 0   & 0   & q_5 & -\theta_{d_2} & 0   & 0\\  
0   & 0   & 0   & 0   & -\theta_{d_2} & -\theta_h   & -\theta_{d_1} & 0   & -\theta_v   & 1   & -\theta_v   & 0   & -\theta_{d_1} & -\theta_h   & -\theta_{d_2} & 0\\
0   & 0   & 0   & 0   & 0   & -\theta_{d_2} & -\theta_h   & -\theta_{d_1} & 0   & -\theta_v   & 1   & -\theta_v   & 0   & -\theta_{d_1} & -\theta_h   & -\theta_{d_2} \\
0   & 0   & 0   & 0   & 0   & 0   & -\theta_{d_2} & q_5 & 0   & 0   & -\theta_v   & q_3 & 0   & 0   & -\theta_{d_1} & q_5\\
0   & 0   & 0   & 0   & 0   & 0   & 0   & 0   & q_5 & -\theta_{d_1} & 0   & 0   & q_1 & q_4 & 0   & 0\\
0   & 0   & 0   & 0   & 0   & 0   & 0   & 0   & -\theta_{d_2}  & -\theta_h  & -\theta_{d_1} & 0   & q_4 & q_2 & q_4 & 0\\
0   & 0   & 0   & 0   & 0   & 0   & 0   & 0   & 0   & -\theta_{d_2} & -\theta_h   & -\theta_{d_1} & 0   & q_4 & q_2 & q_4\\
0   & 0   & 0   & 0   & 0   & 0   & 0   & 0   & 0   & 0   & -\theta_{d_2} & q_5 & 0   & 0   & q_4 & q_1
\end{smallmatrix} \right]
\end{align*}
\caption{\label{eq_Q_mat_4x4} Precision matrix for the 2nd order model shown in Fig. \ref{order2_model}(b), where  $q_1 = 1-(\theta_v+\theta_h+\theta_{b})$,~$q_2 = 1-\theta_h$,~$q_3 = 1-\theta_v$, $q_4 = -(\theta_v+\theta_{b})$, and $q_5 = -(\theta_h+\theta_{b})$.}
\end{figure*}
  
\subsection{Design procedure} \label{design_proc}
We start by generating a training set of GMRF parameters using a large set of sample image blocks. Parameter estimation from image blocks is discussed in Sec. \ref{code_optim_par_est}.  We then use the training set of GMRF parameter vectors to design a codebook (using a vector quantizer design algorithm) for the random parameter vector $\pmb{\theta}$. The parameter vectors in the resulting codebook are used to construct a transform matrix codebook (eigenvectors of precision matrices.) This pre-designed matrix codebook is to be used in an image encoder to pick the best transform matrix for coding each image block.

A standard method for designing a codebook from a training set of vectors is the {\em generalized Lloyd algorithm} (GLA) based on the square-error criterion \cite{Gersho}.  When using this algorithm for GMRF parameter vectors which iterates between the nearest-neighbor and centroid conditions, one has to be careful to ensure that the centroids of the quantization cells in each iteration are also inside the valid parameter space of the GMRF model, i.e., the corresponding precision matrix is positive definite. This would always be the case if the valid parameter space is convex. We have not been able to establish the convexity of the valid parameter space of the model proposed in Sec. \ref{pro_model}. However, the valid parameter spaces of many other finite-lattice homogeneous GMRF models are known to be convex \cite{Balram1,Lakshmanan1,Rue1}. Encouragingly, in all our experiments the GLA always produced parameter vectors that corresponded to positive definite precision matrices, which suggests that the valid parameter space for our GMRF model is possibly (almost) convex.

In most applications, adapting the transform matrix for every individual image block (typically of size $4 \times 4$ or $8\times 8$) can require an impractically large bit rate overhead to signal the transform to the decoder. However, as images and video frames tend to be locally stationary, several adjacent blocks can be coded using the same transform matrix to reduce the signaling overhead. In our experiments, we divide an image into $L \times L$ non-overlapping locally stationary blocks ({\em marcroblocks}) with $L$ suitably chosen and each macroblock is subdivided into $N \times N$ non-overlapping blocks for transform coding ({\em transform blocks}). During the transform coding process, a single transform matrix is used to encode all transform blocks in a given macroblock. The GMRF parameters are thus estimated by assuming all transform blocks in a given macroblock are realizations of the same GMRF. 

\subsection{Transform coding optimized parameter estimation} \label{code_optim_par_est}
Parameter estimation is a constrained optimization problem since the solution for $\pmb{\theta}$ must be such that $\pmb{Q}(\pmb{\theta})$ is  positive definite. The most common method used for GMRF parameter estimation is the ML method \cite{Chellappa3,Rue1}. Let $\tilde{\pmb{C}}$ be the sample covariance matrix of $N \times N$ transform blocks in a given macroblock, and $\pmb{Q}(\pmb{\theta})$ be the precision matrix of the GMRF model for the parameter vector $\pmb{\theta}$. Given the pdf (\ref{gmrf_model}), the log likelihood function  is $J_{\mbox{\scriptsize{ML}}}(\pmb{\theta})=\log\left(\pmb{Q}(\pmb{\theta})\right)-\mbox{Trace}\left(\tilde{\pmb{C}}\pmb{Q}(\pmb{\theta})\right)$. The ML estimate of $\pmb{\theta}$ is given by
\begin{align} \label{ml_estim}
  \pmb{\theta}^*_{\mbox{\scriptsize ML}} & =\arg \max_{\pmb{\theta}} J_{\mbox{\scriptsize{ML}}}(\pmb{\theta}), \mbox{ subject to } \pmb{Q}(\pmb{\theta}) > 0.
\end{align}
However, maximizing likelihood does not necessarily ensure that the resulting GMRFT is optimal for transform coding. We propose here an alternative parameter estimation method which directly minimize the high-rate MSE of transform coding. Let $\pmb{T}(\pmb{\theta})$ be the matrix whose rows are the eigenvectors of $\pmb{Q}(\pmb{\theta})$. The covariance matrix of the transform coefficient vector is $\tilde{\pmb{C}}'(\pmb{\theta})=\pmb{T}(\pmb{\theta})\tilde{\pmb{C}}\pmb{T}(\pmb{\theta})^T$. Under the high-rate assumptions \cite{Gersho}, the minimum MSE of coding an $N \times N$ transform block using the transform matrix $\pmb{T}(\pmb{\theta})$ is given by
\begin{align*}
  J_{TC}(\pmb{\theta})=\frac{\sqrt{3}\pi}{2} K2^{-2R}\left(\prod_{k=1}^K [\tilde{\pmb{C}}'(\pmb{\theta})]_{k,k}\right)^{\frac{1}{K}},
\end{align*}
where $R$ is the bit rate and  $K=N^2$. We find $\pmb{\theta}$ optimized for transform coding by solving
\begin{align}
\pmb{\theta}^*_{\mbox{\scriptsize{TC}}} = \arg \min_{\pmb{\theta}} J_{TC}(\pmb{\theta}),  \mbox{ subject to } \pmb{Q}(\pmb{\theta}) > 0. \label{par_optim}
\end{align}
The problems (\ref{ml_estim}) and (\ref{par_optim}) can be solved using non-linear optimization software, such as {\tt fmincon()} in Matlab. For a given solution, the constraint can be conveniently verified via Cholesky decomposition, see \cite[Sec. 2.7]{Rue1}. It appears difficult to analytically verify  the convexity of either (\ref{ml_estim}) or (\ref{par_optim}). However, in all our experiments, the solutions obtained with many randomly generated initializations produced practically the same parameter estimates in both cases.

It is important to note that the valid parameter space of a finite-lattice non-causal homogeneous GMRF with some given boundary conditions is specific to the lattice size, and therefore parameter estimates based on a given transform block size $N$ can in general  be specific to this block size. This issue is further discussed in the next section. 

\section{Variable block-size GMRFT} \label{sec_scalable_trans}
When coding natural images and video, significant gains can be achieved by spatially adapting the transform block-size based on block structures generated by quad-tree partitioning \cite{Sullivan,Marpe2}. Transform matrices constructed from trigonometric bases such as the 2D-DCT and 2D-DST can be straightforwardly generated on-the-fly for any size. On the other hand, data-driven transforms such as \cite{Sezer2, Boragolla1} are by design, specific to a block-size, and for variable block-size coding, a separate transform matrix codebook must be designed for each target block-size. In contrast, with GMRFT, a single codebook of GMRF-parameters may be used to generate transform matrices of multiple sizes. This is because the GMRF parameters would be identical for all image blocks containing the same texture, regardless of the block-size. By quantizing the space of GMRF parameters, we implicitly quantize the space of image textures.

Suppose we estimate the GMRF parameter vector $\pmb{\theta}$ for a lattice  of size $N\times N$ using the procedure in Sec. \ref{code_optim_par_est} to obtain a positive definite $\pmb{Q}_{K}(\pmb{\theta})$ of size $K \times K$, where $K=N^2$. Suppose we wish to obtain an $K_1 \times K_1$  transform matrix $\pmb{T}_{K_1}(\pmb{\theta})$ for a GMRF defined on a $N_1 \times N_1$ lattice using the same parameter vector $\pmb{\theta}$, where $K_1=N_1^2$. Since any subset of variables in a set of jointly Gaussian variables is also jointly Gaussian, if $\pmb{Q}_{K}(\pmb{\theta})$ is positive definite then  $K_1\times K_1$ precision matrix $\pmb{Q}_{K_{1}}(\pmb{\theta})$ is also guaranteed positive definite for any $K_1 \leq K$. However in general,  $\pmb{Q}_{K_1}(\pmb{\theta})$ is not guaranteed to be positive-definite for $K_1>K$. This is because the valid parameter space of a non-causal homogeneous finite-lattice GMRF with given boundary conditions can depend on the lattice size \cite{Balram1}. One solution is to estimate $\pmb{\theta}$  based on the largest expected transform block-size. This can however greatly reduce the number of image blocks available  in a given training set of images for parameter estimation.

In order to ensure the size-scalability of transform matrices derived from a fixed set of GMRF parameters, we propose to impose a frequently used  sufficient condition that guarantees the positive-definiteness of a $K\times K$ precision matrix $\pmb{Q}(\pmb{\theta})$ regardless of $K$, the {\em diagonal dominance} condition \cite{Balram1,Rue1}
\begin{align}
\pmb{Q}_{ii} > \sum_{j = 1, j\neq i}^K \vert \pmb{Q}_{ij} \vert  , \quad  i=1,\ldots,K. \label{diag_dom}
\end{align}
This requirement can be met by replacing the general constraint $\pmb{Q}(\pmb{\theta})>0$ in (\ref{par_optim}) by (\ref{diag_dom}) in parameter estimation. Furthermore, for homogeneous fields (\ref{diag_dom}) can be reduced to simpler conditions on GMRF parameters. For example, in the case of the  $2$nd-order model in Fig. \ref{order2_model}(b) with $\theta_b=\frac{1}{2}(\theta_{d_1}+\theta_{d_2})$, one can use the triangular inequality $\vert a +b \vert \leq \vert a \vert +\vert b \vert$ to show that (\ref{diag_dom}) is equivalent to
\begin{align*}
 \vert\theta_v\vert + \vert\theta_h\vert + \vert\theta_{d_1}\vert +\vert\theta_{d_2}\vert < \frac{1}{2}
\end{align*}
 regardless of the lattice size. It should be emphasized that the set of $\pmb{\theta}$ that satisfies (\ref{diag_dom}) is often a subset of the parameter space for which  $\pmb{Q}(\pmb{\theta})>0$ for a given matrix size $K$, see for example \cite[Sec V]{Balram1} and \cite[Sec. 2.7.2]{Rue1}. Nonetheless, experimental results  in Sec. \ref{sec_results} show that GMRFTs based on the above model can be very effective.

\section{Experimental Results} \label{sec_results}
 This section presents experimental results demonstrating the advantage of GMRFTs over the 2D-DCT as well as other recently proposed CATs.  For designing GMRFTs, we used macroblocks taken from a training set of 47 gray-scale natural images of various sizes to create a population of GMRF parameter vectors. Parameter estimation for a given macroblock involves estimating the sample covariance matrix of all transform blocks in the macroblock and then applying either the ML method or the  coding optimized method proposed in Sec. \ref{code_optim_par_est}. Each element of the sample covariance matrix was estimated by, first subtracting the average value (which is to be quantized separately) and computing the corresponding pairwise correlations for all pixels within a macroblock (it is reasonable to assume inter-pixel correlations to be approximately spatially invariant over a macroblock, since all transform blocks in the macroblock are assumed to be sampled from the same GMRF.) We considered GMRF models of order 2, 3, and 4 (Fig. \ref{fig:GMRF_NH}(a)) but found that orders higher than 2 resulted in only very marginal improvements in transform coding performance. We here present experimental results obtained with the 2nd-order model shown in Fig. \ref{order2_model} (b) with $\theta_b=\frac{1}{2}(\theta_{d_1}+\theta_{d_2})$. The model is thus restricted to 4 parameters, requiring only 4-dimensional VQ.

In obtaining image coding results, the transform coefficients were quantized using a uniform scalar quantizer whose step-size was chosen to achieve the desired target bit rate. In practice binary entropy coding (e.g. run-length coding, Huffman coding, or arithmetic coding) is applied to quantization indices. In order to avoid the influence of a specific entropy coding scheme on the reported results, we approximated the expected average bit rate by the measured binary entropies of quantization index sequences. The bit rates are reported here in bits per pixel (bpp.)
 
\begin{table}[!bt]
\centering
\caption{\label{tab:Img_blck_EC} Energy compaction (EC) efficiency and coding gain loss relative to the KLT. Results are averages over a large sample of macroblocks.}
\begin{tabular}{|c|c|c|c|}
\hline
  & Parameter estimation  &   & Coding gain \\
Transform & method &  EC (\%)   & loss (dB) \\
\hline\hline
KLT         &  -   & 94.0 &  - \\ 
2D-DCT      &  -       & 86.3   & -4.6  \\
$\mbox{GMRFT}^{\pm}$  & Coding optimized         & 89.8 & -3.6 \\
$\mbox{GMRFT}^{\pm}$ & Maximum likelihood      & 88.6 & -4.0 \\
$\mbox{GMRFT}^{+ }$ & Coding optimized          & 86.8 & -4.5 \\
\hline                                                   
\end{tabular}
\end{table}

\subsubsection{Energy compaction (EC) efficiency of GMRFTs}
Two  metrics commonly used to compare the coding performance of transform matrices are the  EC efficiency \cite[Eq. 3.50]{Clarke} and the {\em coding gain} \cite[Eq. 8.7.1]{Gersho}. For transform coding image blocks, an upper bound and a lower bound for these performance metrics are those for the KLT (which in our case is the eigenvector matrix of the sample covariance matrix) and the 2D-DCT respectively. Table \ref{tab:Img_blck_EC} compares these bounds with the GMRFT performance estimated using  $L=16$ ($16 \times 16$ macroblocks) and $N=8$ ($8\times 8$ transform blocks.) In this case we have defined EC as the fraction of energy in 8 transform coefficients (12.5\% of all 64 coefficients.) Here, $\mbox{GMRFT}^{\pm}$ refers to transforms where the GMRF parameter space is allowed to be real, $\mbox{GMRFT}^{+}$ refers to transforms where the parameter space is constrained to non-negative real (attractive GMRFs.) The table shows the EC efficiency and coding gain of each transform type averaged over 9290 macroblocks (out of a total of 40514) in which $ \mbox{GMRFT}^{\pm}$ transforms designed with coding optimized parameter estimation achieved at least 0.2 dB coding gain over the 2D-DCT. We observed that the coding optimized parameter estimation proposed in Sec. \ref{code_optim_par_est} always resulted in better GMRFTs than the commonly used ML method. Note also that confining the parameter space to an attractive GMRF model results in a loss of transform coding efficiency. We have therefore used $\mbox{GMRFT}^{\pm}$s in obtaining all experimental results presented below.  

\subsubsection{Fixed block-size coding}\label{adapt_code}
\begin{figure}[!tb]
  \centering
  \subfloat[]{\includegraphics[width=0.225\textwidth]{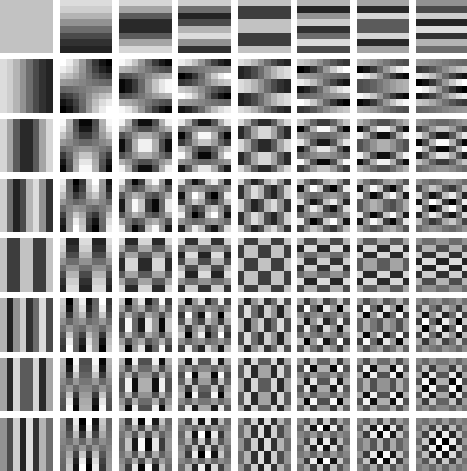}}\hspace{0.15ex}
  \subfloat[]{\includegraphics[width=0.225\textwidth]{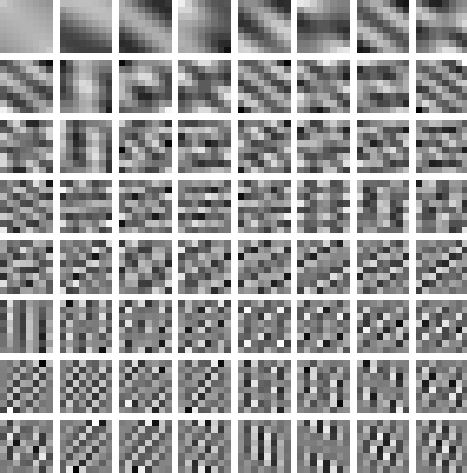}}\\
  \subfloat[]{\includegraphics[width=0.225\textwidth]{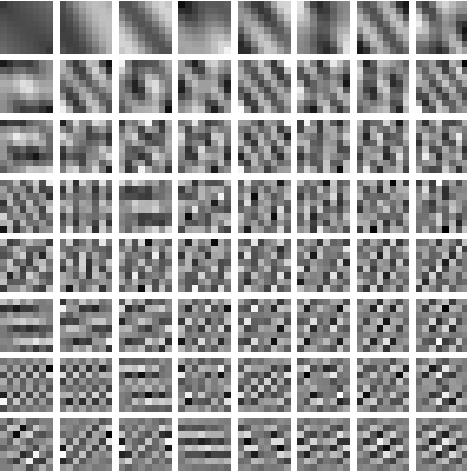}}\hspace{0.15ex}
  \subfloat[]{\includegraphics[width=0.225\textwidth]{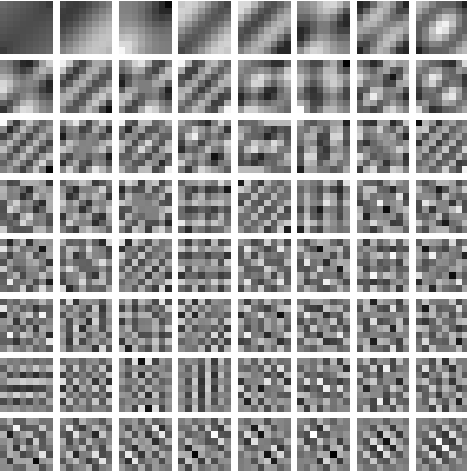}}
  \caption{ \label{fig_GMRFT_CWs} Basis images of  $64 \times 64$ transform matrices:  (a) is 2D-DCT, and (b), (c) and (d) are  3 examples of GMRFTs taken from a codebook containing 7 matrices.}
\end{figure}

Adaptive transform coding requires signaling a quantized GMRFT matrix for each macroblock in an image.  To this end, we considered coding $16 \times 16$ macroblocks in terms of  $8 \times 8$ transform blocks and designed a codebook of 7 GMRF parameter vectors, as described in Sec. \ref{design_proc}. The training set for codebook design was generated as follows.  First we estimated  GMRF parameter vectors from all macroblocks in training images. Then  we pruned the estimated set  by eliminating those parameter vectors whose GMRFTs produced a less than 0.2 dB coding gain over the 2D-DCT on the image blocks from which the parameter vectors were estimated. Upon designing the parameter vector codebook, the GMRFT matrix codebook was generated  by computing the eigenvectors of the precision matrix for each vector in the parameter codebook. The rows of each GMRFT matrix were ordered according to the eigenvalues to ensure energy compaction. Finally we augmented the GMRFT matrix codebook with the 2D-DCT. Fig. \ref{fig_GMRFT_CWs} shows three examples of $64 \times 64$ basis images in GMRFT matrices from the codebook. Note that unlike the basis images in the 2D-DCT which are simply Kronecker products of 1D horizontal and vertical cosine bases, GMRFTs contain non-separable 2D bases matched to diverse textures present in the training image set.  

\begin{figure}[!tb]
  \centering
    \subfloat[{\em Mandril} image ($512\times 512$)]{\includegraphics[width=0.25\textwidth]{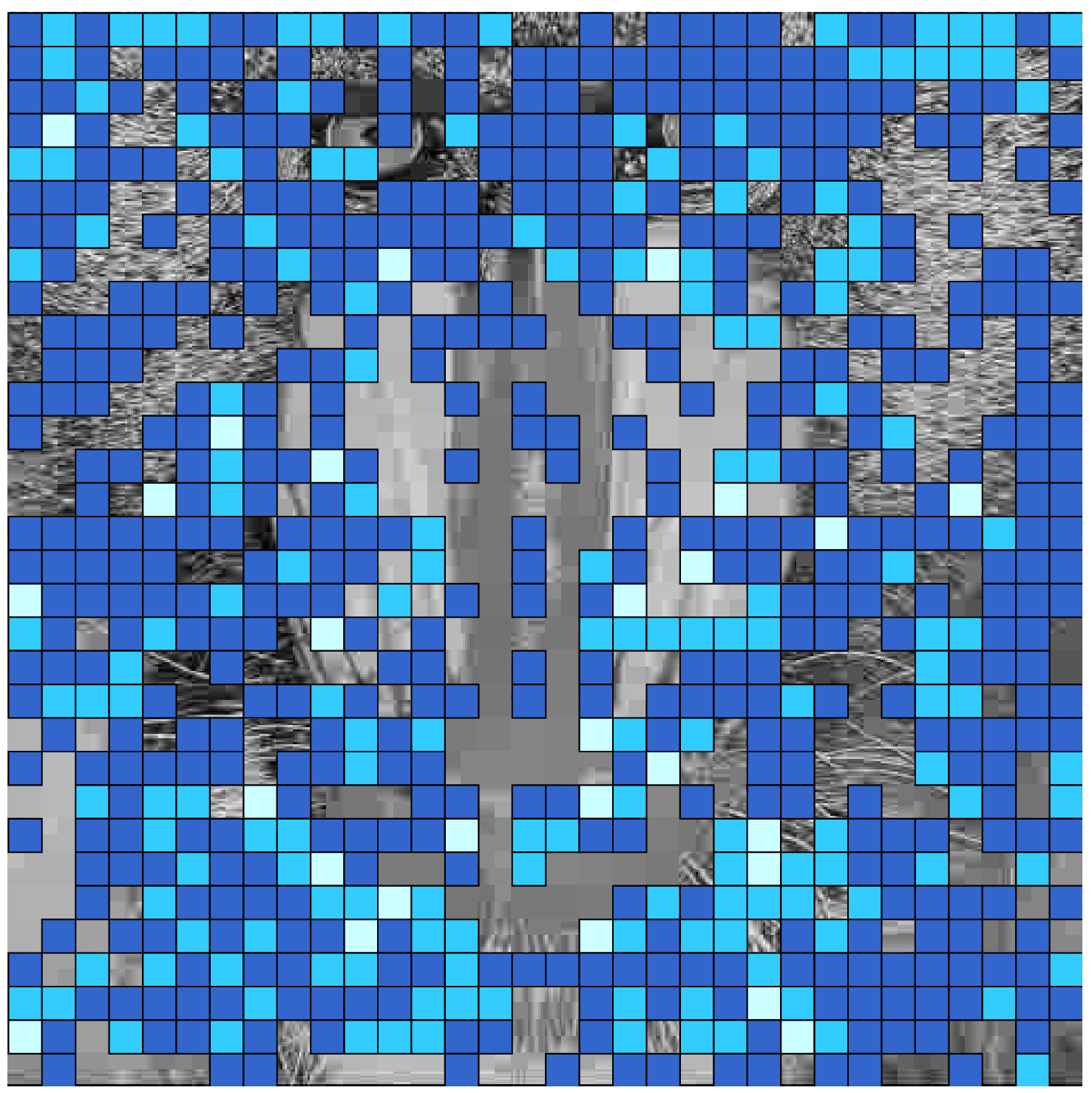}}\\
    \subfloat[{\em Pentagon} image ($1024\times 1024$)]{\includegraphics[width=0.4\textwidth]{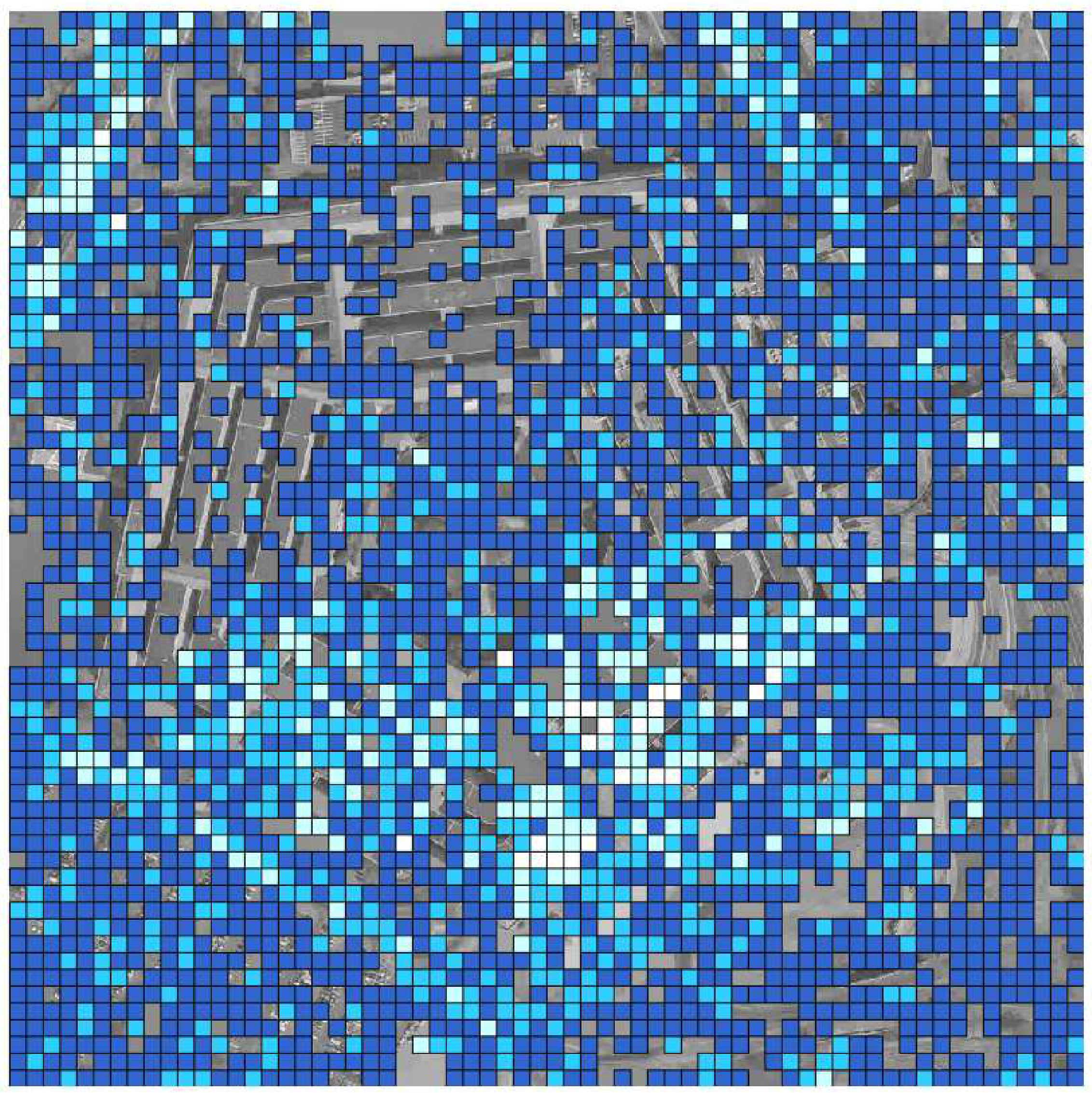}}\\
    \vspace{1ex}
  \includegraphics[width=0.3\textwidth]{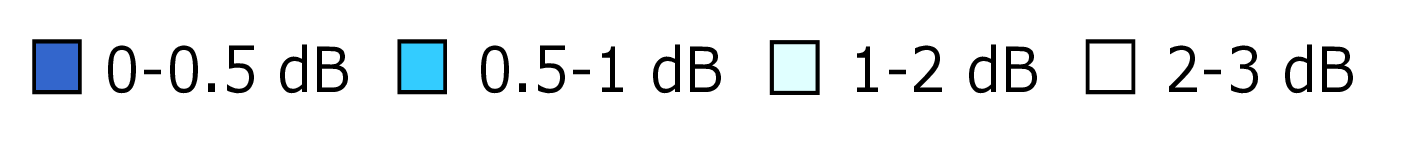}
  \caption{\label{Mandrill_Pentagon} Examples of adaptive transform coding using a GMRFT codebook at 0.4 bpp (see Table \ref{table_Img_PSNR_test_images}.) Codebook size is 8, including the 2D-DCT.  Color coded squares are $16\times 16$ macroblocks coded by a transform other than the 2D-DCT. The color indicates the PSNR gain over the 2D-DCT.}
\end{figure}

\begin{figure}[!tb]
  \centering
  \includegraphics[width=0.4\textwidth]{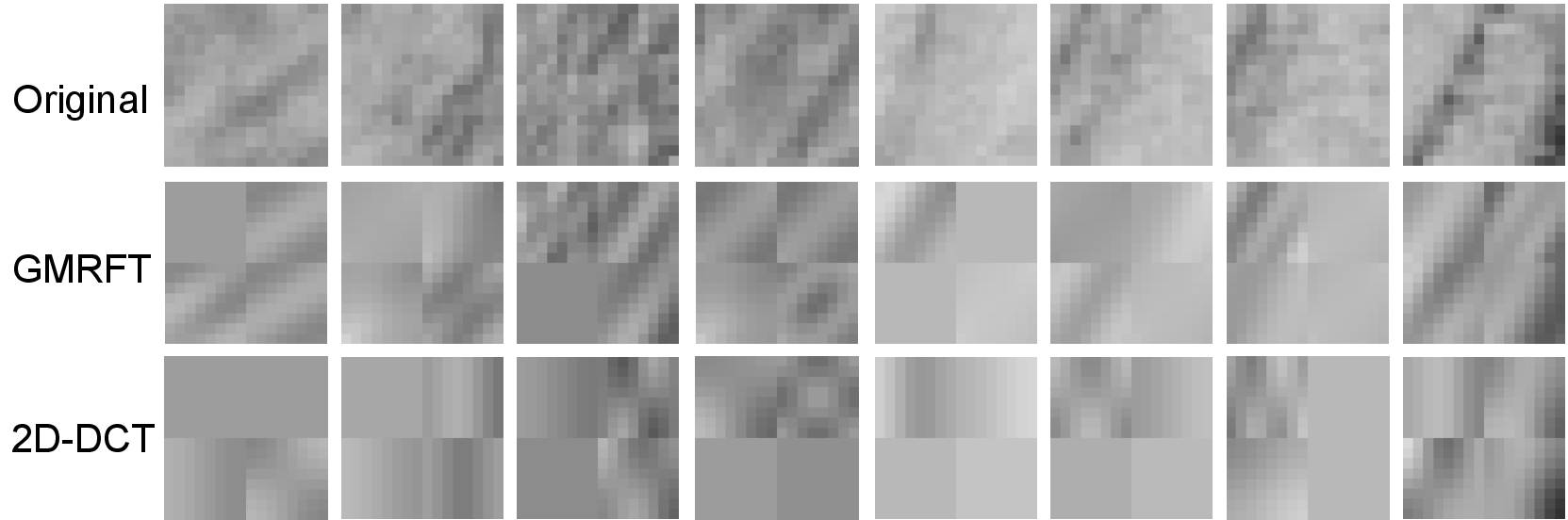}
  \caption{\label{block_examples}Select examples of transform coded $16\times 16$ macroblocks from {\em Manrdrill} image in Fig. \ref{Mandrill_Pentagon}(a), each containing four $8 \times 8$ transform blocks,}
\end{figure}

  We adaptively encoded a set of test images which were not in the training set. Each macroblock  was coded using the transform matrix from the GMRFT matrix codebook which, resulted in the minimum distortion (square error) averaged over all transform blocks in that macroblock. Fig. \ref{Mandrill_Pentagon} shows two examples of adaptively coded images where the color-coded blocks represent the macroblocks for which a transform other than the 2D-DCT got picked from the GMRFT matrix codebook. Fig. \ref{block_examples} shows several examples of coded  macroblocks from the  {\em Mandrill} image, for which a GMRFT outperformed the 2D-DCT. Since the 2D-DCT uses sinusoidal bases in horizontal and vertical directions only, image regions with more complex textures can be better approximated by non-separable GMRFTs learned from actual images, as can be seen in Fig. \ref{block_examples}. Recall also that (see Sec. \ref{sec_model}), 2D-DCT is indeed a special case of GMRFTs which is optimal only for a subset of the overall GMRF parameter space, i.e., parameter vectors corresponding to diagonally symmetric fields. We found that GMRF parameters estimated for many macroblocks in natural images did not satisfy the latter condition.

\begin{figure*}
  \centering
  \includegraphics[width=0.8\textwidth]{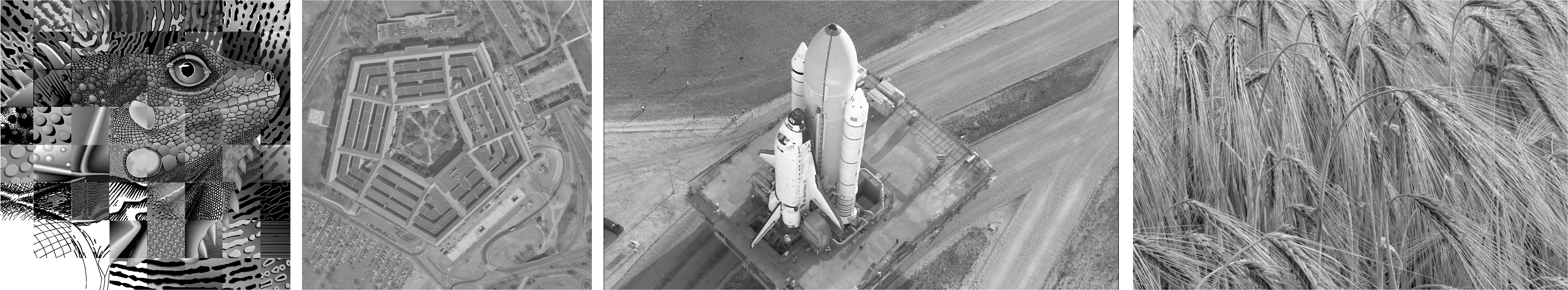}
  \caption{\label{testimages} Test images: (from left to right) {\em Frymire} ($1024 \times 1024$), {\em Pentagon} ($1024 \times 1024)$, {\em Shuttle} ($1280 \times 720$), and {\em Wheat} ($1200 \times 1800$).}
\end{figure*}


\begin{table}
\centering
\caption{\label{table_Img_PSNR_test_images} Image coding PSNR (dB) of KLT, 2D-DCT, and adaptive coding based on a GMRFT matrix codebook of size $8$.}
\resizebox{0.48\textwidth}{!}{
\begin{tabular}{|c|c||c|c|c|c|c|} 
\hline
   &&          &       & \multicolumn{3}{c|}{GMRFT codebook}\\ 
\cline{5-7}
Image         &  Rate  & KLT     & DCT   &           &          & $\%$ non\\
~        & (bpp)   & PSNR    & PSNR  &  PSNR     &  PSNR    & -DCT\\
         &       &         &       &           &  gain$^{*}$& blocks\\
\hline\hline
Babara~  & 0.4   & 34.80 & 30.43 & 30.82   & 0.62                      & 53.6       \\
~        & 0.6   & 37.45 & 32.76 & 33.15   & 0.60                      & 54.3       \\
~        & 0.8   & 39.48 & 34.60 & 34.99   & 0.55                      & 55.3       \\ 
\hline
Lena     & 0.4   & 36.86 & 34.77 & 35.08   & 0.49                      & 60.0       \\
~        & 0.6   & 38.80 & 36.56 & 36.75   & 0.29                      & 63.8       \\
~        & 0.8   & 40.21 & 37.79 & 37.90   & 0.19                      & 66.9       \\ 
\hline
Frymire~ & 0.4   & 23.60 & 21.13 & 21.38   & 0.40                      & 60.0       \\
~        & 0.6   & 25.92 & 22.77 & 23.04   & 0.43                      & 62.6       \\
~        & 0.8   & 27.96 & 24.24 & 24.54   & 0.44                      & 65.4       \\ 
\hline
Pirate   & 0.4   & 33.66 & 32.03 & 32.20   & 0.28                      & 63.9       \\
~        & 0.6   & 35.66 & 33.69 & 33.82   & 0.23                      & 66.4       \\
~        & 0.8   & 37.15 & 34.97 & 35.00   & 0.11                      & 68.6       \\ 
\hline
Pentagon & 0.4   & 31.46 & 29.31 & 29.61   & 0.41                      & 72.1       \\
~        & 0.6   & 33.26 & 30.66 & 30.91   & 0.35                      & 71.0       \\
~        & 0.8   & 34.64 & 31.79 & 31.98   & 0.28                      & 72.3       \\ 
\hline
Mandrill & 0.4   & 26.10 & 24.48 & 24.74   & 0.37                      & 67.3       \\
~        & 0.6   & 27.88 & 25.85 & 26.13   & 0.39                      & 68.7       \\
~        & 0.8   & 29.45 & 27.08 & 27.28   & 0.30                      & 70.1       \\
\hline
\multicolumn{7}{l}{$^{*}$PSNR gain over the 2D-DCT in only non-DCT macroblocks.}
\end{tabular}
}
\end{table}

\begin{figure}
  \centering
  \includegraphics[width=0.495\textwidth]{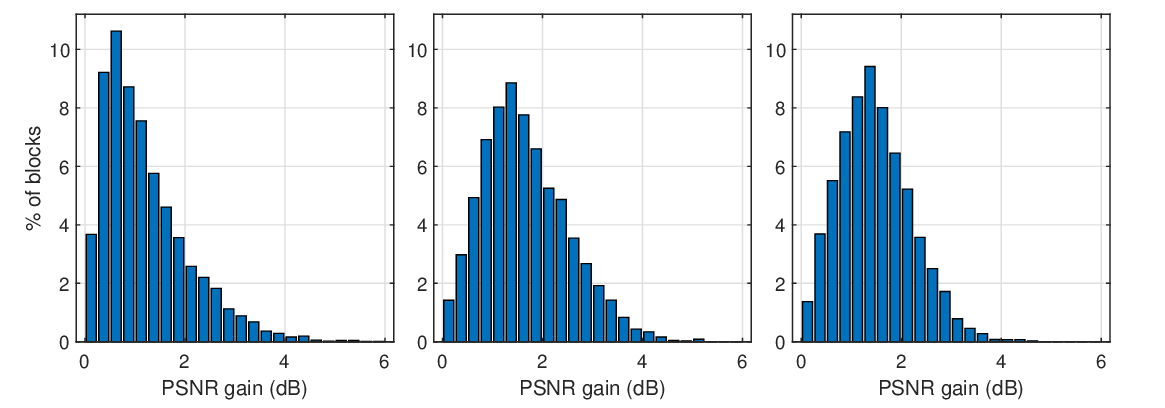}
  \caption{\label{PSNR_histo} Histograms of PSNR gains (over the 2D-DCT) in GMRFT-coded macroblocks at bit rates 0.4 bpp (left), 0.6 bpp (center), and 0.8 bpp.}
\end{figure}
    
Table \ref{table_Img_PSNR_test_images} compares the peak-signal-to-noise ratio (PSNR) of GMRFT-based adaptive coding  with the sample-KLT upper bound and the DCT lower bound, for a selection of test images (some of which are shown in Fig. \ref{testimages}.) The column {\em PSNR gain} shows the average PSNR gain (over the 2D-DCT) in only those macroblocks which  got coded with a transform other than the 2D-DCT, while the column {\em \% non-DCT blocks} shows the percentage of such  macroblocks (see for example Fig. \ref{Mandrill_Pentagon}.) These two columns show the  advantage of using GMRFTs better matched to the texture characteristics of each macroblock, as also evident from Fig. \ref{block_examples}. In order to illustrate this more clearly, we present in Fig. \ref{PSNR_histo} histograms of PSNR gains in macroblocks from a test set of 14 images, which got coded using a transform other than the 2D-DCT (compared to the PSNR if the same macroblocks were coded using 2D-DCT). While the average PSNR gains in Table \ref{table_Img_PSNR_test_images} are mostly in 0.3-0.6 dB range, GMRFTs achieve gains up to 6 dB for some macroblocks in test images. Note that the bit rates of GMRFT-based coding shown in this table also include the additional bit rate required to signal the transform matrix to the decoder on a per macroblock basis. This bit rate overhead was observed to be only about 1-3\% of the total bit rate in all cases for a transform matrix codebook of size 8.
 
\subsubsection{Variable block-size adaptive coding}
As already noted in Sec. \ref{sec_scalable_trans}, quad-tree based variable block-size coding based on the 2D-DCT has been observed in previous work to yield considerable improvement in coding gain with natural images. In order to apply GMRFT matrix codebooks for variable block-size coding, we used a {\em base codebook} of GMRF parameters designed for $8 \times 8$ transform blocks as in Sec. \ref{adapt_code} above, which can be essentially viewed as a codebook of parameterized image textures. In our experiments we used the diagonal dominance condition as discussed in Sec. \ref{sec_scalable_trans} to ensure that GMRF parameters are block-size independent.  Given a set of GMRF parameters from the base codebook, a transform for an image block of any size having the same texture can be  generated in a straightforward manner by setting-up the corresponding precision matrix and computing its eigenvectors.

\begin{figure}[!bt]
  \centering
\includegraphics[width=0.4\textwidth]{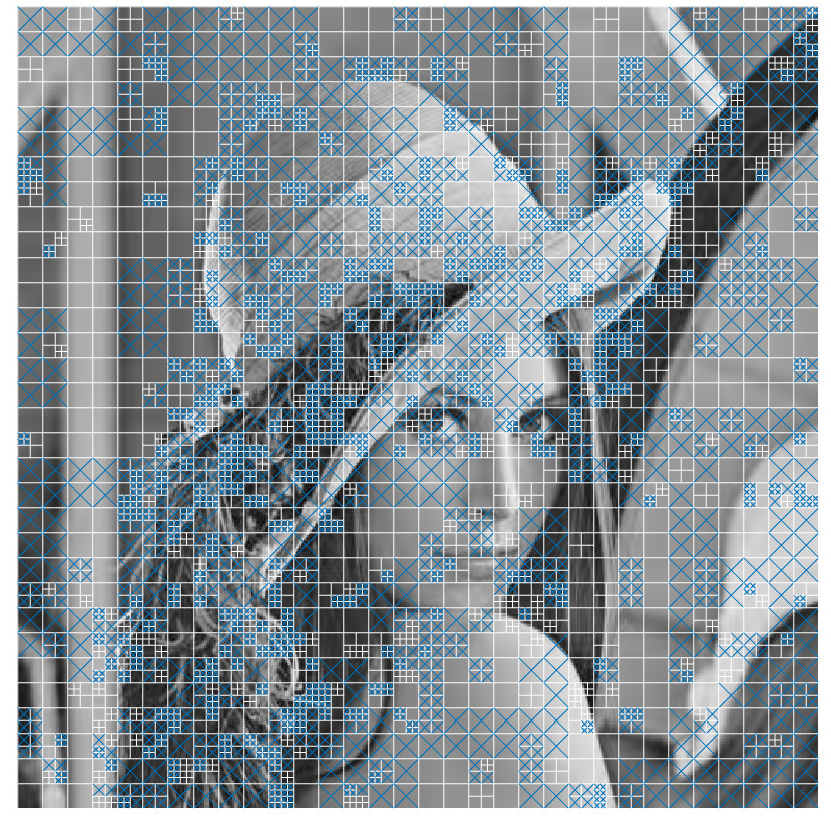}
\caption{ \label{fig_Lena_CB_QT_3S} GMRFT-based variable block-size transform coding of $512 \times 512$ Lena image  at 0.4 bpp based on 3-stage quad-tree partitioning of $32\times 32$ macroblocks. Transform codebook also included 7 GMRFTs and the 2D-DCT. Blocks marked with crosses are those coded with a GMRFT.} 
\end{figure}

In our experiments, we applied top-down quad-tree decomposition \cite{Sullivan} to $L \times L$ macroblocks in the image to be coded. In the first decomposition stage,  a macroblock is subdivided into four $\frac{L}{2} \times \frac{L}{2}$ transform blocks. In subsequent stages, each  $N \times N$ transform block is further subdivided into four $\frac{N}{2} \times \frac{N}{2}$, if coding each of these transform blocks using the best GMRFT resulted in a reduction of the total square-error. Note that 4 transform blocks in this case can use different transforms. The procedure is recursively applied to each transform block until a specified minimum block size $N_{\min}\times N_{\min}$ is reached. A target bit rate was achieved by repeating the procedure for different quantization step-sizes (a more systematic approach is  decomposition based on a rate-distortion measure \cite{Sullivan}.) Experimental results presented here have been obtained with a transform matrix codebook containing 7 GMRFTs and the 2D-DCT.  The estimate for the total bit rate is the sum of the entropies of the quantized transform coefficients, the transform matrix indices, and a bit sequence defining the quad-tree structure.  Fig. \ref{fig_Lena_CB_QT_3S} shows the quad-tree structure obtained by adaptive coding the $512 \times 512$ {\em Lena} image at the bit rate 0.4 bpp, using $L=32$ and $N_{\min}=4$, i.e., a 3 stage quad-tree decomposition. 

\begin{table*}[t]
  \centering
  \caption{\label{quad_tree1} PSNR (dB) comparison of fixed block-size coding and variable block-size coding. In all cases, macroblock size is $32 \times 32$.  {\em Fixed block-size} refers to coding each macroblock using $8 \times 8$ transform blocks and a common transform matrix. {\em Single stage} refers to coding each macroblock using $16 \times 16$ transform blocks and a different (best)  transform matrix for each transform block.}
\resizebox{\textwidth}{!}{
\begin{tabular}{|c|c||c|c|c|c|c||c|c|c||c|c|c||c|c|c|} 
\hline
\multirow{3}{*}{Image} & Rate  & \multicolumn{5}{c||}{Fixed macroblcok size} & \multicolumn{9}{c|}{Quad-tree partition}                                                          \\ 
\cline{8-16}
                       & (bpp) & \multicolumn{5}{c||}{(32x32)}               & \multicolumn{3}{c||}{Single stage} & \multicolumn{3}{c||}{2-stage~} & \multicolumn{3}{c|}{3-stage}  \\ 
\cline{3-16}
                       &       & DCT   & GMRFT & DDCT \cite{Zeng} & QOT \cite{Boragolla1}   & SOT\cite{Sezer2}        & DCT   & GMRFT & DDCT             & DCT   & GMRFT & DDCT        & DCT   & GMRFT & DDCT          \\ 
\hline\hline
Lena                   & 0.4   & 34.63 & 34.86 & 34.73 & 34.8  & 34.76      & 35.42 & 35.61 & 35.47             & 35.75 & 36.21 & 35.97         & 35.75 & 36.24 & 35.91         \\
($512\times 512$)              & 0.6   & 36.48 & 36.59 & 36.58 & 36.6  & 36.59      & 36.97 & 37.13 & 37.06             & 37.41 & 37.70 & 37.60         & 37.42 & 37.88 & 37.71         \\
~                      & 0.8   & 37.78 & 37.89 & 37.79 & 37.87 & 37.8       & 38.13 & 38.19 & 38.16             & 38.55 & 38.87 & 38.68         & 38.55 & 39.03 & 38.89         \\ 
\hline
Babara                 & 0.4   & 30.13 & 30.44 & 30.28 & 30.75 & 30.52      & 31.4  & 31.65 & 31.56             & 31.55 & 31.89 & 31.65         & 31.54 & 31.92 & 31.66         \\
($512\times 512$)              & 0.6   & 32.55 & 32.85 & 32.7  & 33.25 & 32.97      & 33.65 & 33.95 & 33.78             & 33.78 & 34.12 & 33.88         & 33.78 & 34.14 & 33.90         \\
~                      & 0.8   & 34.48 & 34.76 & 34.6  & 35.11 & 34.82      & 35.49 & 35.68 & 35.64             & 35.65 & 36.02 & 35.66         & 35.67 & 36.02 & 35.75         \\ 
\hline
Camaraman              & 0.4   & 37.65 & 37.69 & 37.78 & 37.72 & 37.75      & 38.38 & 38.59 & 38.82             & 38.67 & 39.00 & 38.85         & 38.68 & 38.94 & 38.81         \\
($512\times 512$)              & 0.6   & 40.25 & 40.22 & 40.44 & 40.31 & 40.31      & 40.98 & 41.15 & 41.31             & 41.35 & 41.39 & 41.32         & 41.22 & 41.49 & 41.34         \\
~                      & 0.8   & 42.34 & 42.31 & 42.36 & 42.36 & 42.36      & 43.09 & 43.27 & 43.35             & 43.27 & 43.26 & 43.42         & 43.35 & 43.36 & 43.39         \\ 
\hline
Mandrill               & 0.4   & 24.37 & 24.59 & 24.41 & 24.59 & 24.56      & 24.68 & 24.95 & 24.74             & 24.85 & 25.20 & 24.86         & 24.90 & 25.31 & 24.96         \\
($512\times 512$)              & 0.6   & 25.8  & 26.02 & 25.84 & 26.04 & 25.98      & 26.06 & 26.34 & 26.11             & 26.21 & 26.60 & 26.22         & 26.25 & 26.75 & 26.38         \\
~                      & 0.8   & 27.04 & 27.21 & 27.07 & 27.26 & 27.19      & 27.29 & 27.53 & 27.33             & 27.46 & 27.88 & 27.50         & 27.57 & 28.06 & 27.67         \\ 
\hline
Pirate                 & 0.4   & 31.88 & 31.94 & 31.87 & 31.86 & 31.89      & 32.24 & 32.41 & 32.29             & 32.73 & 33.07 & 32.83         & 32.76 & 33.20 & 32.97         \\
($1024\times 1024$)            & 0.6   & 33.59 & 33.66 & 33.61 & 33.59 & 33.59      & 33.85 & 33.93 & 33.84             & 34.38 & 34.58 & 34.48         & 34.45 & 34.79 & 34.64         \\
~                      & 0.8   & 34.9  & 34.91 & 34.87 & 34.89 & 34.89      & 35.08 & 35.15 & 35.01             & 35.53 & 35.83 & 35.72         & 35.65 & 36.09 & 35.91         \\ 
\hline
Pentagon~              & 0.4   & 29.2  & 29.42 & 29.26 & 29.42 & 29.37      & 29.4  & 29.57 & 29.45             & 29.79 & 30.20 & 29.97         & 29.95 & 30.51 & 30.24         \\
($1024\times 1024$)            & 0.6   & 30.63 & 30.77 & 30.64 & 30.82 & 30.79      & 30.69 & 30.81 & 30.7              & 31.12 & 31.51 & 31.35         & 31.34 & 31.94 & 31.74         \\
~                      & 0.8   & 31.79 & 31.89 & 31.79 & 31.94 & 31.98      & 31.77 & 31.85 & 31.8              & 32.22 & 32.61 & 32.45         & 32.46 & 33.00 & 32.87         \\ 
\hline
Wheat~                 & 0.4   & 24.66 & 25.43 & 24.97 & 25.61 & 25.48      & 25.34 & 26.14 & 25.62             & 25.36 & 26.25 & 25.58         & 25.37 & 26.28 & 25.61         \\
($1200\times 1800$)            & 0.6   & 26.65 & 27.54 & 26.98 & 27.79 & 27.56      & 27.36 & 28.21 & 27.66             & 27.40 & 28.40 & 27.66         & 27.40 & 28.47 & 27.71         \\
~                      & 0.8   & 28.4  & 29.28 & 28.75 & 29.63 & 29.37      & 29.14 & 30.03 & 29.43             & 29.16 & 30.27 & 29.47         & 29.18 & 30.30 & 29.52         \\ 
\hline
Shuttle~               & 0.4   & 35.38 & 35.53 & 35.39 & 35.41 & 35.41      & 35.96 & 36.08 & 35.94             & 36.37 & 36.64 & 36.36         & 36.64 & 36.73 & 36.44         \\
($1280\times 720$)             & 0.6   & 38.04 & 38.04 & 38.04 & 38.05 & 38.05      & 38.53 & 38.61 & 38.55             & 39.07 & 39.26 & 39.01         & 39.34 & 39.41 & 39.22         \\
~                      & 0.8   & 40.16 & 40.11 & 40.06 & 40.16 & 40.17      & 40.66 & 40.72 & 40.78             & 41.22 & 41.43 & 41.24         & 41.49 & 41.74 & 41.52         \\ 
\hline
Frymire                & 0.4   & 20.66 & 20.82 & 20.72 & 20.82 & 20.92      & 20.76 & 20.9  & 20.81             & 21.65 & 22.50 & 22.22         & 22.36 & 23.66 & 23.68         \\
($1024\times 1024$)              & 0.6   & 22.4  & 22.55 & 22.48 & 22.6  & 22.84      & 22.21 & 22.36 & 22.29             & 23.52 & 24.66 & 24.52         & 24.50 & 26.85 & 26.88         \\
~                      & 0.8   & 23.92 & 24.03 & 24.06 & 24.17 & 24.51      & 23.51 & 23.69 & 23.65             & 25.03 & 26.57 & 26.52         & 26.29 & 29.03 & 29.28         \\
\hline
\end{tabular}}
\end{table*}

\begin{table}
  \centering
  \caption{\label{quad_tree2} PSNR (dB) gains of GMRFT-based variable block-size coding over the 2D-DCT and DDCT counterparts, and fixed block-size coding with QOT and SOT.}
  \resizebox{0.48\textwidth}{!}{
\begin{tabular}{|c|c||c|c||c|c|} 
\hline
\multirow{3}{*}{Image} & Rate  & DCT         & DDCT \cite{Zeng}        & \multicolumn{2}{c|}{Fixed block-size}  \\
                       & (bpp) & (3-stage    & (3-stage    & \multicolumn{2}{c|}{~}                  \\ 
\cline{5-6}
                       & ~     & quad-tree) & quad-tree) & QOT \cite{Boragolla1} & SOT \cite{Sezer2}                             \\ 
\hline \hline
Lena                   & 0.4   & 0.49        & 0.33        & 1.44 & 1.48                             \\
                       & 0.6   & 0.46        & 0.17        & 1.28 & 1.29                             \\
~                      & 0.8   & 0.47        & 0.13        & 1.16 & 1.23                             \\ 
\hline
Babara                 & 0.4   & 0.38        & 0.26        & 1.17 & 1.40                             \\
                       & 0.6   & 0.35        & 0.24        & 0.89 & 1.17                             \\
~                      & 0.8   & 0.36        & 0.28        & 0.91 & 1.20                             \\ 
\hline
Cameraman              & 0.4   & 0.26        & 0.13        & 1.22 & 1.19                             \\
                       & 0.6   & 0.27        & 0.16        & 1.18 & 1.18                             \\
~                      & 0.8   & 0.01        & 0           & 1.00 & 1.00                             \\ 
\hline
Mandrill               & 0.4   & 0.41        & 0.35        & 0.72 & 0.75                             \\
                       & 0.6   & 0.50        & 0.37        & 0.71 & 0.77                             \\
~                      & 0.8   & 0.50        & 0.40        & 0.80 & 0.87                             \\ 
\hline
Pirate~                & 0.4   & 0.44        & 0.23        & 1.34 & 1.31                             \\
                       & 0.6   & 0.34        & 0.15        & 1.20 & 1.20                             \\
~                      & 0.8   & 0.44        & 0.18        & 1.20 & 1.20                             \\ 
\hline
Pentagon~              & 0.4   & 0.56        & 0.27        & 1.09 & 1.14                             \\
                       & 0.6   & 0.59        & 0.19        & 1.12 & 1.15                             \\
~                      & 0.8   & 0.54        & 0.13        & 1.06 & 1.02                             \\ 
\hline
Wheat~                 & 0.4   & 0.92        & 0.67        & 0.67 & 0.80                             \\
                       & 0.6   & 1.07        & 0.76        & 0.68 & 0.91                             \\
~                      & 0.8   & 1.12        & 0.78        & 0.67 & 0.93                             \\ 
\hline
Shuttle~               & 0.4   & 0.31        & 0.29        & 1.32 & 1.32                             \\
                       & 0.6   & 0.24        & 0.19        & 1.36 & 1.36                             \\
~                      & 0.8   & 0.39        & 0.23        & 1.58 & 1.57                             \\ 
\hline
Frymair                & 0.4   & 1.30        & 0       & 2.84 & 2.74                             \\
                       & 0.6   & 2.35        & 0       & 4.25 & 4.01                             \\
~                      & 0.8   & 2.74        & -0.25       & 4.86 & 4.52                             \\
\hline
\end{tabular}}
\end{table}


In Table \ref{quad_tree1} we compare the PSNR performance of variable block-size coding and fixed block-size coding. All results have been obtained with $32 \times 32$ macroblocks. For fixed block-size coding, a transform block-size of $8 \times 8$ was used. In addition to the standard 2D-DCT, we also show the PSNRs of fixed block-size coding  with the DDCT \cite{Zeng} as well as  two other recently reported CAT design methods, sparse orthonormal transforms (SOT) \cite{Sezer2} and  quantization-optimized transforms (QOT) \cite{Boragolla1}. Experimental results in \cite{Sezer2,Boragolla1} show that SOT and QOT outperform those designed with many other previously reported methods. However, both SOT and QOT designs are specific to a transform block-size, which is cumbersome in the case of variable block-size coding. DDCT matrices can on the other hand be easily constructed for any transform block size. The results in Table \ref{quad_tree1} show that for fixed block-size coding GMRFT performs close to SOT and QOT regardless of the fact that, unlike GMRFTs which are model-based transforms involving only 4 parameters, both QOT and SOT are orthonormal  matrices defined by $N^2(N^2-1)/2$ free parameters, all of which are learnt from training data. More importantly Table \ref{quad_tree1} shows that, with variable block-size coding, GMRFT provides substantial PSNR improvement over fixed block-size coding. In order to more clearly see the achievable PSNR improvements, in Table  \ref{quad_tree2}, we summarize the PSNR gains of GMRFT-based variable block-size (3-stage quad-tree)  coding relative to the 2D-DCT and DDCT counterparts, and fixed block-size coding with SOT and QOT. Note that, in a few cases DDCT performs comparably or better than the choice available in the GMRFT matrix codebook. This is because, for strongly directional textures DDCT can be quite good. GMRFTs on the other hand are capable of capturing more complex textures, and hence in most cases GMRFTs outperform the DDCT even in variable block-size coding. 

\section{Conclusions} \label{sec_conclude}
With adaptive transform coding in mind, we investigated an approach to quantizing local KLT matrices of image blocks in a 4-dimensional space independent of the matrix size. This approach is based on representing image textures by a 4-parameter GMRF model. We also showed that the required GMRF parameters can be estimated from sample image data to maximize the transform coding gain. One attractive feature of the proposed GMRFTs is the ability to use a single codebook of GMRF parameter vectors to generate transform matrices of arbitrary sizes which has applications in variable block-size adaptive transform coding. Our experiments with quad-tree based variable block-size image coding have shown that the proposed GMRFTs  could outperform the 2D-DCT by as much as 2.7 dB as well as the DDCT by up to 0.8 dB for some images.

Another potential application of GMRFTs is {\em shape-adaptive} transform coding. Our preliminary results indicate that GMRFTs can outperform the shape-adaptive DCT \cite{Sikora} and graph-transforms considered in some previous work \cite{Fracastoro}.

\bibliographystyle{IEEEbib}
\bibliography{IEEEabrv,gmrft_v4}
\end{document}